\documentclass[10pt,twocolumn,letterpaper]{IEEEtran}
\usepackage{amsmath,epsfig}
\usepackage{algorithm}
\usepackage{bm}
\usepackage{tikz}
\usepackage{graphicx}
\usetikzlibrary{positioning}
\usetikzlibrary{shapes.multipart}
\usepackage{pgfplots}
\usepackage[usenames,dvipsnames]{pstricks}
\usepackage{epsfig}
\usepackage{pst-grad} 
\usepackage{pst-plot} 
\usepackage{subfigure}
\usepackage[capitalize]{cleveref}

\usepackage{subfigure}
\usepackage[T1]{fontenc}
\usepackage{amsmath}
\usepackage{graphics}
\include{psfig}
\usepackage{epsfig}
\usepackage{array}
\usepackage{psfrag}
\usepackage{amssymb}
\usepackage{color}
\usepackage{pstricks,pst-node,pst-text,pst-3d,pst-plot}
\usepackage{cite}

\usepackage{ulem}
\definecolor{dgreen}{rgb}{0, 0.8, 0.1}

\begin{document}

\title{3D Near-Field Millimeter-Wave Synthetic Aperture Radar Imaging}
\author{Shahrokh Hamidi$^*$\thanks{Shahrokh Hamidi is with the Faculty of Electrical and Computer
Engineering, University of Waterloo, 200 University Ave W, Waterloo, ON., Canada, N2L 3G1.
e-mail: \texttt{Shahrokh.Hamidi@uwaterloo.ca}.} \and and Safieddin Safavi-Naeini$^*$\thanks{Safieddin Safavi-Naeini is with the Faculty of Electrical and Computer Engineering, University of Waterloo, 200 University Ave W, Waterloo, ON., Canada, N2L 3G1.
e-mail: \texttt{
safavi@uwaterloo.ca}.}}

\maketitle

\begin{tikzpicture}[remember picture, overlay]
      \node[font=\small] at ([yshift=-1cm]current page.north)  {This paper has been accepted for publication in the IEEE $\rm 19^{th}$ International
Symposium on Antenna Technology and Applied Electromagnetics, 2021. \copyright IEEE};
\end{tikzpicture}

\begin{abstract}
In this paper, we present 3D high resolution radar imaging process at millimeter-Wave (mmWave) frequencies by creating the effect of a large aperture synthetically. We use a low-cost fully integrated Frequency Modulated Continuous Wave (FMCW) radar operating at $\rm 79\;GHz$ and then perform Synthetic Aperture Radar (SAR) imaging in the near-filed zone.

At the end, we conduct a real experiment and present the reconstructed image.
\end{abstract}

\begin{IEEEkeywords}
Synthetic Aperture Radar,  mmWave FMCW radar, Near-Filed Imaging
\end{IEEEkeywords}

\section{introduction}
Synthetic Aperture Radar (SAR) is a very well developed subject to achieve high angular resolution by creating the effect of a large aperture synthetically \cite{Cumming, Soumekh, shahrokh}. The synthetic aperture can be created linearly \cite{Cumming, Soumekh} or circularly \cite{Soumekh, Soumekh_2}. In the former case, the data collection can be performed in the stripmap \cite{Cumming} or spotlight \cite{Spotlight, Soumekh} mode. Recently, the possibility of SAR imaging at millimeter-Wave (mmWave) frequencies have been investigated mostly in stripmap mode and promising results have been generated \cite{MMW_SAR, Yanik_2D, Yanik_2D_SAR_MIMO, Yanik_3D_SAR_MIMO, Yanik_3D_SAR_MIMO_}. As the center frequency increases the lateral resolution obtained by the SAR system also enhances. As a result, by operating at mmWave frequencies higher lateral resolution can be achieved.

In \cite{MMW_SAR_Zhuge}, 3D mmWave SAR imaging at near-field has been presented by combining the SAR system with a Multiple Input Multiple Output (MIMO) arrays.

In \cite{MMW_SAR}, the idea of compressed sensing has been exploited to further increase the resolution of the image. The sparsity has been enforced using $l_1$ norm. They have cast the imaging process as an optimization problem and have added an extra term related to the total variation method to remove the effect of the artifacts and shadows.

In \cite{MMW_SAR_Topology}, 3D mmWave MIMO-SAR imaging at near-field has been presented by considering arbitrary topologies for the MIMO array.

In \cite{Yanik_2D, Yanik_2D_SAR_MIMO, Yanik_3D_SAR_MIMO, Yanik_3D_SAR_MIMO_}, the authors consider 3D mmWave SAR imaging using fully integrated MIMO FMCW radar systems.  In contrast to \cite{Yanik_2D}, where only one transmit and one receive antenna have been used, in \cite{Yanik_2D_SAR_MIMO, Yanik_3D_SAR_MIMO, Yanik_3D_SAR_MIMO_} the MIMO mode of the radars have been integrated into the imaging system. By exploiting the MIMO structure in the vertical direction the authors in \cite{Yanik_2D_SAR_MIMO, Yanik_3D_SAR_MIMO, Yanik_3D_SAR_MIMO_} have considerably reduced the time that it takes to collect the data over the synthetic aperture.

In this paper, we follow \cite{MMW_SAR_Zhuge, MMW_SAR, MMW_SAR_Topology, Yanik_2D, Yanik_2D_SAR_MIMO, Yanik_3D_SAR_MIMO, Yanik_3D_SAR_MIMO_} and present 3D high resolution SAR imaging at near-field. We then use the AWR1243 FMCW radar from Texas Instruments (TI) for the data collection as well as a low cost 2D stepper motor to create the synthetic aperture. Subsequently, we perform a real experiment and reconstruct the image based on the method which we present in the paper.

The rest of the paper has been organized as follows. In Section \ref{System Model}, we present the system model and formulate the imaging problem. Section \ref{Image Reconstruction} discusses the image reconstruction algorithm. Finally, Section \ref{Experimental Results} presents our experimental results which is followed by concluding remarks.
\section{System Model}\label{System Model}
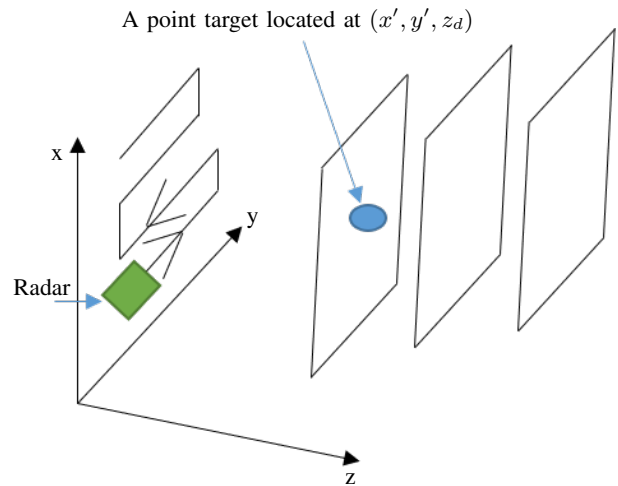
\begin{figure}[htb]
\centering
\begin{tikzpicture}
  \node (img1)  {\includegraphics[scale=1]{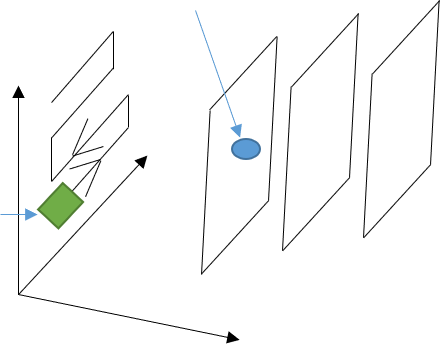}};
  \node[above=of img1, node distance=0cm, xshift=0.2cm, yshift=-7.3cm,font=\color{black}] {\small z};
  \node[above=of img1, node distance=0cm, xshift=-0.5cm, yshift=-1.3cm,font=\color{black}] {\small A point target located at $(x^{\prime},y^{\prime},z_d)$};
 \node[above=of img1, node distance=0cm, xshift=-1.1cm, yshift=-3.9cm,font=\color{black}] {\small y};
   \node[above=of img1, node distance=0cm, xshift=-3.9cm, yshift=-4.8cm,font=\color{black}] {\small Radar};
   \node[above=of img1, node distance=0cm, xshift=-3.7cm, yshift=-3cm,font=\color{black}] {\small x};
\end{tikzpicture}
\caption{The geometry of the model.
\label{fig:Model_Geometry}}
\end{figure}

In this section, we describe the system model. Fig.~\ref{fig:Model_Geometry} illustrates the geometry of the imaging problem.
The received signal from a point reflector located at $(x^{\prime}, y^{\prime}, z_d)$ is described as
\begin{align}
\label{beat_signal}
s_0(k,x,y) = &p(x^{\prime},y^{\prime},z_d)\times \nonumber \\ &e^{\displaystyle -j2k\sqrt{(x-x^{\prime})^2+(y-y^{\prime})^2+z^2_d}}.
\end{align}
In (\ref{beat_signal}), $k = \frac{4\pi}{c}(f_0+\beta t)$, in which, $\beta = \frac{b}{T}$ where $b$ and $T$ stand for the bandwidth and the chirp length, respectively. The complex parameter $p(x^{\prime},y^{\prime},z_d)$ is the reflection coefficient of the point reflector located at $(x^{\prime},y^{\prime},z_d)$. Furthermore, $f_0$ is the start frequency and $c$ is the speed of light in vacuum.

The FMCW radar we consider in this paper is a multi-static radar meaning that the transmit and receive antennas are separated. However, if $d \ll \sqrt{4\alpha \frac{c}{f_c} \mathcal{R}}$, in which $\mathcal{R}$ is measured with respect to the midpoint of the line that connects the transmit and receive antennas \cite{Bi_Mono}, $d$ is the distance between the transmit and receive antennas and $f_c$ is the center frequency, we can then perform our analysis based on the mono-static assumption. In fact, that is exactly what we have done in (\ref{beat_signal}), where the radial distance is with respect to the midpoint of the line that connects the transmit and receive antennas.

The radar is mounted on a tow-axis motorized scanner which spans the x-y plane to create the 2D synthetic aperture. The aperture plane is at $z=0$. Next, we consider a continuum of point reflectors on the plane located at $z_d$. Based on the superposition principle, the reflection from all the point reflectors can then be added that results in
\begin{align}
\label{beat_signal_}
& s(k,x,y) = \int\int p(x^{\prime},y^{\prime},z_d) \times \nonumber \\ & e^{\displaystyle -j2k\sqrt{(x-x^{\prime})^2+(y-y^{\prime})^2+{z^2_d}}}dx^{\prime}dy^{\prime}.
\end{align}
We can rewrite (\ref{beat_signal_}) as
\begin{align}
\label{beat_signal_conv}
s(k,x,y) =  p(x,y,z_d)\ast_x\ast_ye^{\displaystyle -j2k\sqrt{x^2+y^2+z^2_d}},
\end{align}
where $\ast_x$ and $\ast_y$ represent the convolution operation in the x and y directions, respectively.
From (\ref{beat_signal_conv}) it is clear that $e^{\displaystyle -j2k\sqrt{x^2+y^2+z^2_d}}$ is the impulse response of the 2D SAR system.
Upon taking a 2D spatial Fourier transform with respect to the x and y directions from (\ref{beat_signal_conv}), based on the principle of stationary phase (POSP) \cite{Optics}, we obtain
\begin{align}
\label{beat_signal_2dfft}
S(k,k_x,k_y) = & P(k,k_x,k_y)\times \nonumber \\
&e^{\displaystyle j\left(z_d \sqrt{4k^2 - k^2_x - k^2_y}-x^{\prime}k_x-y^{\prime}k_y \right)},
\end{align}
where $P(k,k_x,k_y) = \mathfrak{F}_{2D}\{p(k,x,y)\}$ in which $\mathfrak{F}_{2D}$ stands for 2D Fourier transform in the $x$ and $y$ directions.
\section{Image Reconstruction}\label{Image Reconstruction}
The goal in SAR imaging is to estimate the complex reflectivity coefficient field of the targets. We can obtain the complex reflectivity coefficient $p(x,y,z_d)$ from (\ref{beat_signal_2dfft}) as
\begin{align}
\label{RMA_Algorithm_f}
&\hat{p}(x,y,z_d) = \nonumber \\
&\int  \mathfrak{F}^{-1}_{2D}\left (S(k,k_x,k_y)e^{\displaystyle - j z_d \sqrt{4k^2 - k^2_x - k^2_y}}\right)dk,
\end{align}
where $\mathfrak{F}^{-1}_{2D}$ stands for 2D inverse Fourier transform in the $x$ and $y$ directions and $\hat{p}(x,y,z_d)$ is the estimated value for the complex reflectivity coefficient of the targets  on the plane that has been located at $z_d$.

As can be seen from (\ref{RMA_Algorithm_f}), we first cancel the effect of the phase term $e^{\displaystyle - j z_d \sqrt{4k^2 - k^2_x - k^2_y}}$ per each given value of $z_d$. We will then take a 2D inverse Fourier transform in the x and y directions. Following that, we sum over the different values of $k$ as they all contain the energy of the target on the $z_d$ plane. By choosing different values for $z_d$ we can reconstruct the image of targets located at different distances from the radar and this way we can obtain a 3D image.

\section{Experimental Results}\label{Experimental Results}
\begin{figure}
\psfrag{x [mm]}[][]{$x$ [mm]}
\centerline{
\includegraphics[height=3.5cm,width=4.5cm]{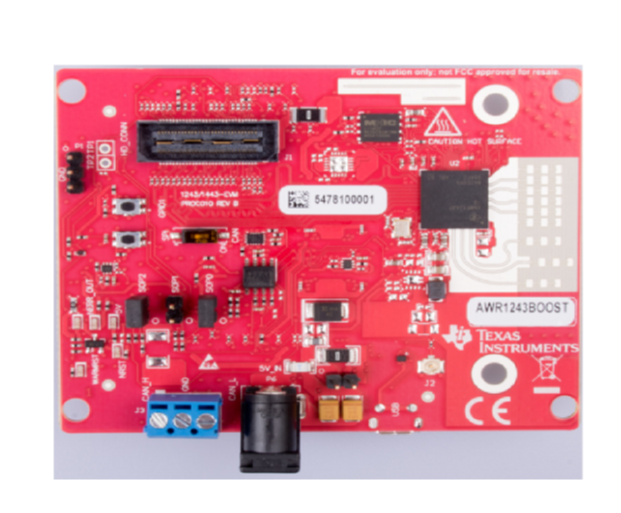}
\hspace*{0.1in}
\includegraphics[height=3.5cm,width=4cm]{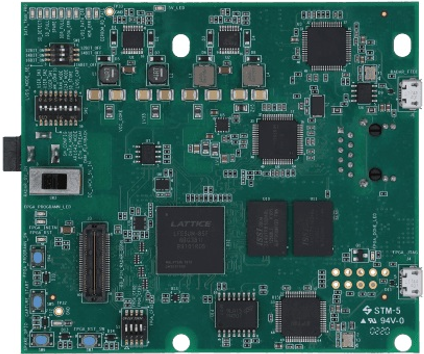}
}
\centerline{(a)\hspace*{4.5cm}(b)}
\vspace*{0.2in}
\caption{a) the AWR1243 radar from TI, b) the DCA1000EVM real time data-capture board.
\label{fig:TI}}
\end{figure}

\begin{figure}[htb]
\centering
\begin{tikzpicture}
  \node (img1)  {\includegraphics[scale=0.5]{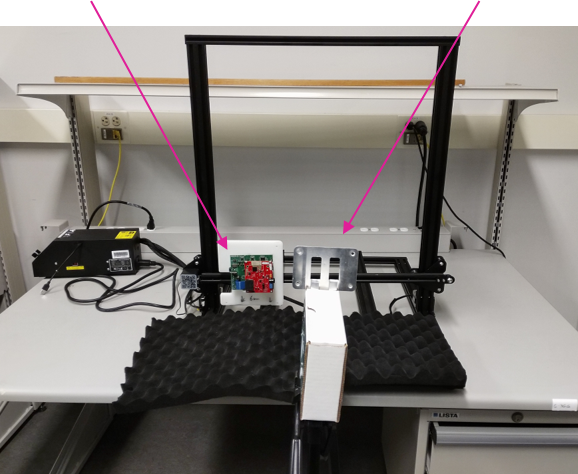}};
  \node[above=of img1, node distance=0cm, xshift=-1.6cm, yshift=-1.2cm,font=\color{black}] {\small The radar};
   \node[above=of img1, node distance=0cm, xshift=1.6cm, yshift=-1.2cm,font=\color{black}] {\small The sample under test};
  \node[above=of img1, node distance=0cm, xshift=0cm, yshift=-5.8cm,font=\color{black}] {(a)};
   \node[below  = 0.5cm of img1](img2){\includegraphics[scale=0.5]{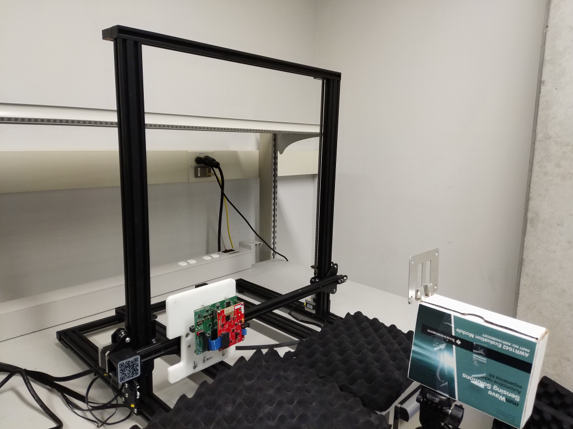}};
    \node[below=of img2, node distance=0cm, xshift=0cm, yshift=0.8cm,font=\color{black}] {(b)};
\end{tikzpicture}
\caption{a)  the test setup which shows the radar (the RF and the FPGA boards), the sample under test and the 2D aperture over which the synthetic aperture is built, b) another view of the test setup.
\label{fig:Setup}}
\end{figure}
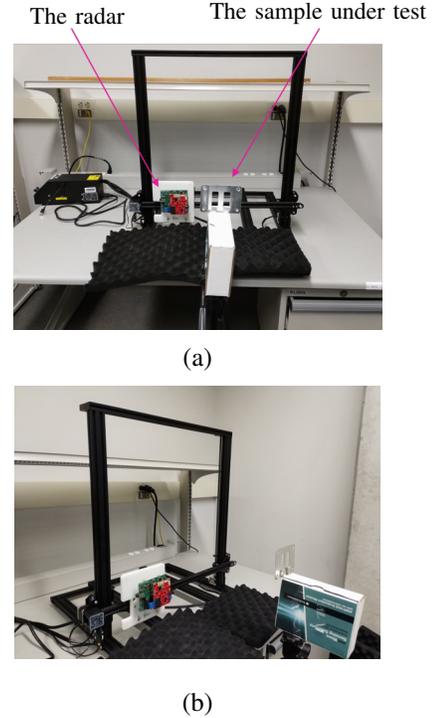

In this section, we present our experimental results.
We have used the AWR1243 FMCW radar as well as the DCA1000EVM real-time data-capture board from TI which have been shown in Fig.~\ref{fig:TI}. We have only used one TX and one RX for data collection.
The test setup has been illustrated in Fig.~\ref{fig:Setup}.

The start frequency and the bandwidth of the radar have been set to  $\rm 77\;GHz$ and $\rm 3.84\;GHz$, respectively. The length of the chirp signal is $\rm 60 \; \mu s$. The samples in the $x$ direction have been taken every $\rm 0.5\; mm$. In the $y$ direction the sample spacing is $\rm 2\; mm$. We have taken $596$ samples in the horizontal direction and $69$ samples in the vertical direction. As a result we have created a $69 \times 596$ elements array.

Therefore, the size of the synthetic array is $\rm 13.8^{cm} \times 29.8^{cm}$ which results in $\rm z_{Fresnel} = 5.41\;m$ for the Fresnel distance and since the target has been located at $\rm z_d = 30\;cm$ from the synthetic aperture, thus the target is inside the Fresnel zone and consequently the near-field imaging is required.

\begin{figure}[htb]
\centering
\begin{tikzpicture}
  \node (img1)  {\includegraphics[scale=0.7]{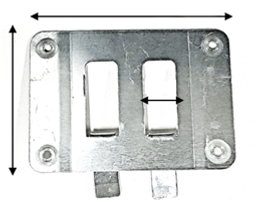}};
  \node[above=of img1, node distance=0cm, xshift=0cm, yshift=-1.5cm,font=\color{black}] {\small $\rm 10.5 \; cm$};
  \node[above=of img1, node distance=0cm, xshift=0.7cm, yshift=-3cm,font=\color{black}] {\tiny $\rm 2 \; cm$};
   \node[above=of img1, node distance=0cm, xshift=-2.6cm, yshift=-3cm,font=\color{black}] {\small $\rm 7 \; cm$};
    \node[above=of img1, node distance=0cm, xshift=0cm, yshift=-5.5cm,font=\color{black}] {(a)};
   \node[below  = 0.5cm of img1](img2){\includegraphics[scale=0.7]{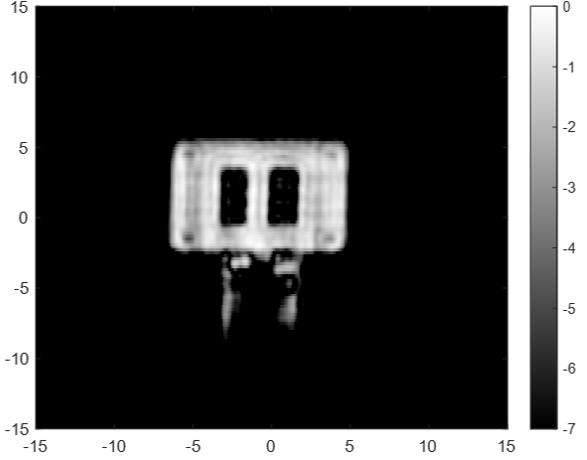}};
   \node[below=of img2, node distance=0cm, xshift=-0.3cm, yshift=1.3cm,font=\color{black}] {\small $\rm x[cm]$};
   \node[below=of img2, node distance=0cm, xshift=-3.9cm, yshift=4.5cm,font=\color{black}] {\small $\rm y[cm]$};
    \node[below=of img2, node distance=0cm, xshift=0cm, yshift=0.8cm,font=\color{black}] {(b)};
\end{tikzpicture}
\caption{a)  the test sample, b) the reconstructed image.
\label{fig:Exp1}}
\end{figure}
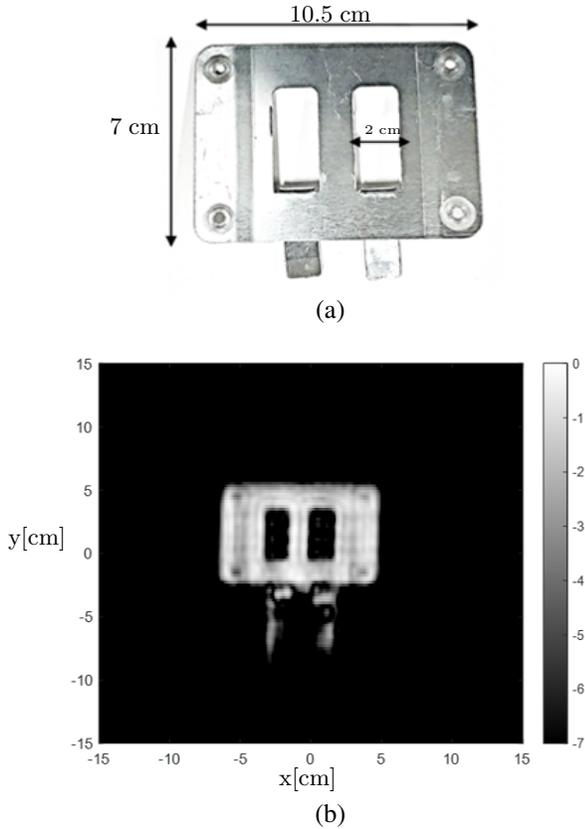

The test sample which is a metallic object has been illustrated in Fig.~\ref{fig:Exp1}-(a).
Fig.~\ref{fig:Exp1}-(b) depicts the result of applying (\ref{RMA_Algorithm_f}) to the data collected from the test sample shown in Fig.~\ref{fig:Exp1}-(a) based on the test setup depicted in Fig.~\ref{fig:Setup}.
As can be seen from Fig.~\ref{fig:Exp1}-(b), a high resolution image has been created from the test sample using the SAR imaging technique.
\section{Conclusion}
In this paper, we presented 3D near-field high resolution mmWave SAR imaging technique. We described the model and the image reconstruction method and at the end applied the algorithm to the experimental data gathered from a FMCW radar operating at $\rm 79\;GHz$ and showed the final result.

\bibliographystyle{IEEEtran}
\bibliography{Biblio}

\end{document}